\documentclass[prl,twocolumn,showpacs,preprintnumbers,amsmath,amssymb]{revtex4-1}

\usepackage{graphicx}
\usepackage{dcolumn}
\usepackage{bm}
\usepackage{color}

\begin{document}


\title{The temperature dependence of FeRh's transport properties}

\author{S.~Mankovsky, S.~Polesya, K.~Chadova  and  H.~Ebert}
\affiliation{Department Chemie,
  Ludwig-Maximilians-Universit\"at M\"unchen, 81377 M\"unchen, Germany}%
\author{ J. B.~Staunton }
\affiliation{Department of Physics, University of Warwick, Coventry, UK}%
\author{T.~Gruenbaum, M. A. W.~Schoen and C. H.~Back }
\affiliation{Department of Physics, Regensburg University, Regensburg, Germany}%
\author{ X. Z. Chen, C. Song }
\affiliation{Key Laboratory of Advanced Materials (MOE), School of Materials Science and Engineering, Tsinghua University, Beijing 100084, China.}%

\date{\today}
             
\begin{abstract} 
The finite-temperature transport properties of FeRh compounds are
investigated by first-principles Density Functional Theory-based calculations. 
The focus is on the behavior of the longitudinal
resistivity with rising temperature, which exhibits an abrupt decrease at the
metamagnetic transition point, $T = T_m$ between ferro- and antiferromagnetic phases.
A detailed electronic structure investigation for $T \geq 0$~K explains this feature 
and demonstrates the important role of (i) the difference of
the electronic structure at the Fermi level between the two magnetically
ordered states and (ii) the different degree of thermally induced magnetic disorder in
the vicinity of $T_m$, giving different contributions to the
resistivity. 
To support these conclusions, we also describe
the temperature dependence of the spin-orbit induced
anomalous Hall resistivity and Gilbert damping parameter.
For the various response quantities considered the impact of
thermal lattice vibrations and spin fluctuations on their temperature dependence is investigated
in detail. Comparison with corresponding experimental data finds in
general a very good agreement. 
\end{abstract}

\pacs{Valid PACS appear here}
\maketitle


For a long time the ordered equiatomic FeRh alloy has attracted much attention
owing to its intriguing temperature dependent magnetic and magnetotransport 
properties. The crux of these features of this CsCl-structured material is the first order transition from
an antiferromagnetic (AFM) to ferromagnetic (FM) state when the
temperature is increased above $T_m = 320$~K \cite{KH62,BB95a}.  In this context
the drop of the electrical resistivity that is observed  across the
metamagnetic transition is of central interest. Furthermore, if the AFM to FM 
transition is induced by an applied magnetic field,
a pronounced magnetoresistance (MR) effect is found
experimentally with a measured MR ratio $\sim 50 \%$ at room temperature 
\cite{AIM+95,BB95a,dVLM+13}. The temperature of the metamagnetic
transition 
as well as the  MR ratio can be tuned by addition of small amounts of
impurities \cite{Wal64,Kou66,BB95a,BJL13,LNS09}. 
These properties make FeRh-based materials very attractive for future
applications in data storage devices. 
The origin of the large MR effect in FeRh, however, is still under debate.
Suzuki et al. \cite{SNI+11} suggest that, for deposited thin FeRh films,
the main mechanism stems from the spin-dependent scattering of
 conducting electrons on localized  magnetic moments
 associated with  partially occupied electronic $d$-states \cite{Kas95} at grain boundaries.
 Kobayashi et 
 al. \cite{KMA01} have also discussed the MR effect in the bulk ordered
 FeRh system 
 attributing its origin to the modification of the
 Fermi surface across the metamagnetic transition. 
 So far only one theoretical investigation of the MR effect in FeRh has
 been carried out on an ab-initio level \cite{KDT15}.  

\medskip

The present study is based on spin-polarized, electronic
structure calculations using the fully relativistic multiple scattering
KKR (Korringa-Kohn-Rostoker)
 Green function method \cite{SPR-KKR6.3,EKM11,SM}.
This approach allowed to calculate 
the transport properties of  FeRh  at finite temperatures
on the basis of 
 the linear response formalism using the Kubo-St\v{r}eda 
expression for the conductivity
tensor \cite{Str82,LKE10}
%
\begin{eqnarray}
\label{eq:sig}
\sigma_{\rm \mu\nu}
&
=
&
\frac{\hbar }{4\pi N\Omega}
{\rm Trace}\,\big\langle \hat{j}_{\rm \mu} (G^+(E_F)-G^-(E_F))
\hat{j}_{\rm \nu} G^-(E_F)
\nonumber
\\
&&
\qquad \qquad \quad
-  \hat{j}_{\rm \mu} G^+(E_F)\hat{j}_{\rm \nu}(G^+(E_F)-G^-(E_F))\big\rangle_{\rm c}
\label{eq:bru}
\; ,
\end{eqnarray} 
%
where $\Omega$ is the volume of the unit cell, $N$ is the number of
sites, $\hat{j}_{\mu}$ is the relativistic current operator and 
$G^\pm(E_F)$ are the electronic retarded and advanced Green functions,
respectively, calculated at the Fermi energy $E_F$.
In Eq. (\ref{eq:bru}) the orbital current term has been omitted 
 as it only provides small corrections to the
prevailing contribution arising from the first term in the 
case of a cubic metallic system~\cite{NHK10, LGK+11, TKD12}.

Here we focus on the finite temperature transport
properties of FeRh. In order to take into account
electron-phonon and electron-magnon scattering effects in the
calculations, the so-called alloy analogy model \cite{EMKK11,EMC+15a} is
used. Within this approach the temperature induced spin (local moment) and lattice
excitations are treated as localized, slowly varying degrees of freedom with
temperature dependent amplitudes. 
Using the adiabatic approximation in the calculations of transport
properties, and accounting for the random character 
of the motions, the evaluation of the thermal average
over the spin and lattice excitations in Eq.~(\ref{eq:bru}) is 
reduced to a calculation of the configurational average over the
local lattice distortions and magnetic moment orientations,
$\langle ... \rangle_{\rm c}$,
using the recently reported approach \cite{EMKK11,EMC+15a} 
which is based on the 
coherent potential approximation (CPA)
alloy theory  \cite{Vel69,But85,TKD+02b}.

\begin{figure}[h]
\includegraphics[width=0.4\textwidth,angle=0,clip]{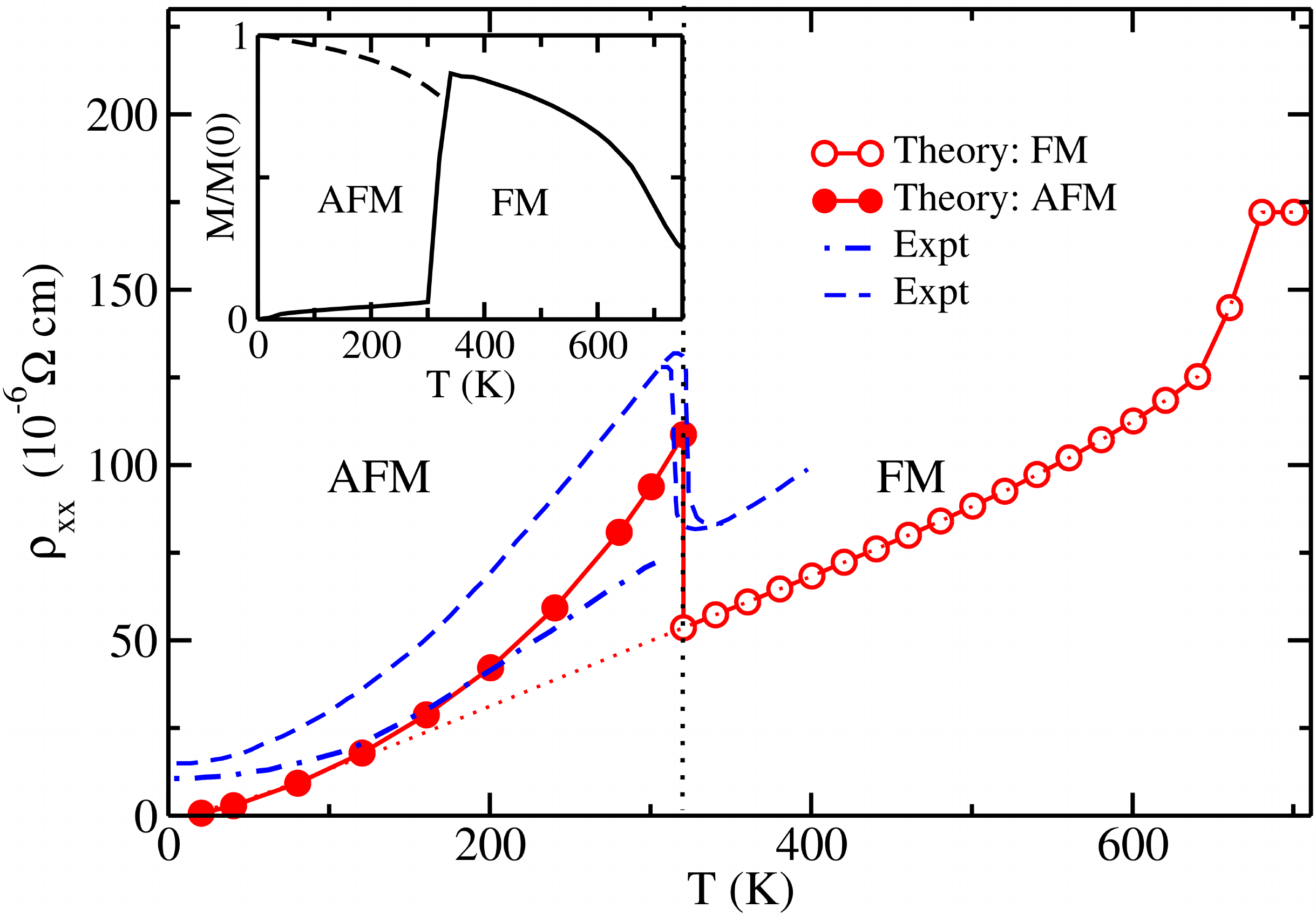}\;(a)\\ 
\includegraphics[width=0.4\textwidth,angle=0,clip]{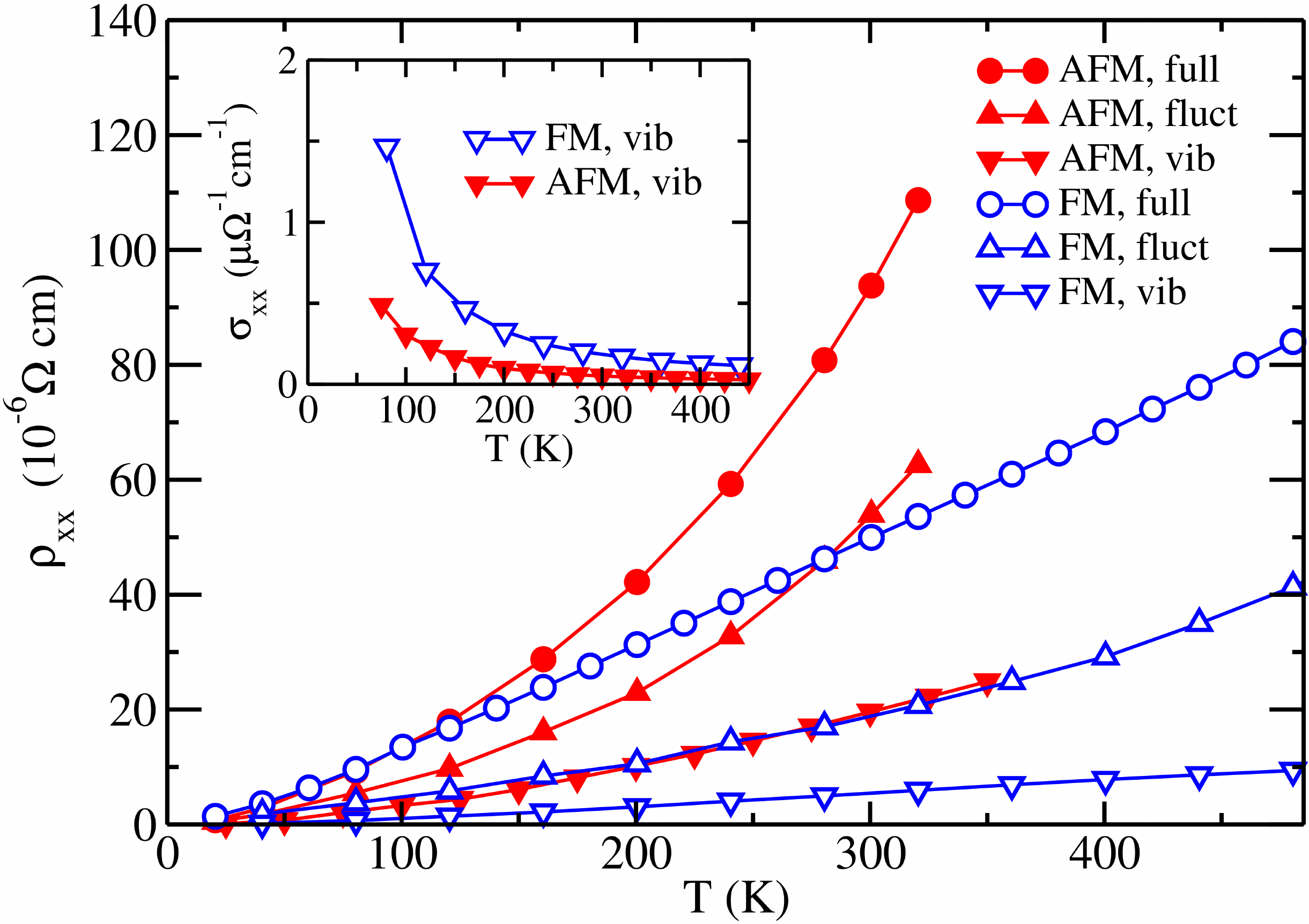}\;(b)
\caption{\label{fig:FeRh_Rho} (a) Calculated longitudinal resistivity
  (closed circles - AFM state, open 
  circles - FM state) in comparison with 
  experiment \cite{BB95a}. The dashed line represents the results for
  Fe$_{0.49}$Rh$_{0.51}$, while the dash-dotted line gives results for
  (Fe-Ni)$_{0.49}$Rh$_{0.51}$ with the Ni concentration $x=0.05$ to
  stabilize the FM state at low temperature).
  The inset represents the relative magnetization of a Fe
  sub-lattice as a function of temperature obtained from MC simulations.
  (b) electrical resistivity
  calculated for the AFM (closed symbols) and FM (open symbols) states
  accounting for all thermal scattering effects (circles) as well as
  effects of lattice vibrations (diamond) and spin fluctuations
  (squares) separately. The inset shows the temperature dependent
  longitudinal conductivity for the AFM and FM states due to lattice
  vibrations only.
 }  
\end{figure}

To account for the effect of spin fluctuations, which we 
describe in a similar way as is done within the disordered 
local moment (DLM) theory \cite{GPS+85}, the angular
distribution of thermal spin moment fluctuations 
is calculated using the results of Monte Carlo (MC)
simulations. These are based on ab-initio  exchange coupling parameters
 and reproduce the finite temperature magnetic properties for
the AFM and FM state in both the low- ($T < T_m$) and high-temperature ($T >
T_m$) regions very well \cite{PMK+16}. 
Figure~\ref{fig:FeRh_Rho}(a), inset, shows the temperature dependent 
magnetization, $M(T)$, for one of the two Fe sublattices aligned
antiparallel/parallel to each other in the AFM/FM state, calculated across 
the temperature region covering both AFM and FM states of the system.  
%
The 
differing behavior of the magnetic order $M(T)$ in the two phases has important
consequences for the transport properties as discussed below.

Figure \ref{fig:FeRh_Rho}(a) shows the calculated electrical 
resistivity as a function of temperature, $\rho_{xx}(T)$, accounting for the effects of
electron scattering from thermal spin and lattice excitations, and compares it
with experimental data. There is clearly a rather good theory-experiment agreement
especially concerning the difference  
$\rho_{xx}^{AFM}(T_m) - \rho_{xx}^{FM}(T_m)$ at the AFM/FM transition, $T_m=$ 320K.
The AFM state's resistivity increases more steeply with
temperature when compared to that of the FM state, 
that has also been calculated for temperatures below the metamagnetic
transition temperature (dotted line). 
Note that the experimental measurements have been performed for a
sample with $1\%$ intermixing between the Rh and Fe sublattices leading to a
finite residual resistivity at $T \to 0$~K, and as a consequence there is a 
shift of the experimental $\rho_{xx}(T)$ curve with respect to the
theoretical one \cite{SBS+14}. 

We can separate out the contributions of spin fluctuations and lattice vibrations
to the electrical resistivities, $\rho_{xx}^{fluc}(T)$ and
$\rho_{xx}^{vib}(T)$, respectively. 
These two components have been calculated for finite temperatures keeping the
atomic positions undistorted to find $\rho_{xx}^{fluc}(T)$ and fixed collinear
orientations of all magnetic moments to find $\rho_{xx}^{vib}(T)$,
respectively. The results for 
the AFM and FM states are shown in Fig. \ref{fig:FeRh_Rho}(b), 
where again the FM (AFM) state has also been considered below
  (above) the transition temperature $T_m$.
For both magnetic states the local moment fluctuations have a
dominant impact on the resistivity. One can also see that both
components, $\rho_{xx}^{fluc}(T)$ and  $\rho_{xx}^{vib}(T)$, in the AFM
state have a steeper increase with temperature than those of the FM state.

The origin of this behavior can be clarified by referring to   
Mott's model \cite{Mot64} with its distinction between delocalized
$sp$-electrons, which primarily determine the transport properties owing to their
high mobility, and the more localized $d$-electrons. Accordingly, the conductivity
should depend essentially on (see. e.g. \cite{TPM12}): (i) the carrier (essentially 
$sp$-character) concentration $n$ and (ii) the relaxation time
$\tau \sim [V_{scatt}^2n(E_F)]^{-1}$, where 
$V_{scatt}$ is the average scattering potential and $n(E_F)$ the total density of
states at the Fermi level. This model has been used, in
particular, for qualitative discussions of the origin of the GMR effect 
in heterostructures consisting of magnetic layers separated by non-magnetic spacers. 
In this case the GMR effect can be attributed to the spin dependent
scattering of conduction electrons which leads to a
dependence of the resistivities on the relative orientation of magnetic layers, 
parallel or antiparallel, 
assuming the electronic structure of non-magnetic spacer to be unchanged. 
These arguments, however, cannot be straightforwardly applied to CsCl-structured
FeRh, even though it can be pictured as a layered system with one atom thick layers,
since the electronic structure of FeRh shows strong modifications across the
AFM-FM transition as discussed, for example, by Kobayashi et al. \cite{KMA01} to 
explain the large MR effect in FeRh.

 \begin{figure}[h]
 \includegraphics[width=0.23\textwidth,angle=0,clip]{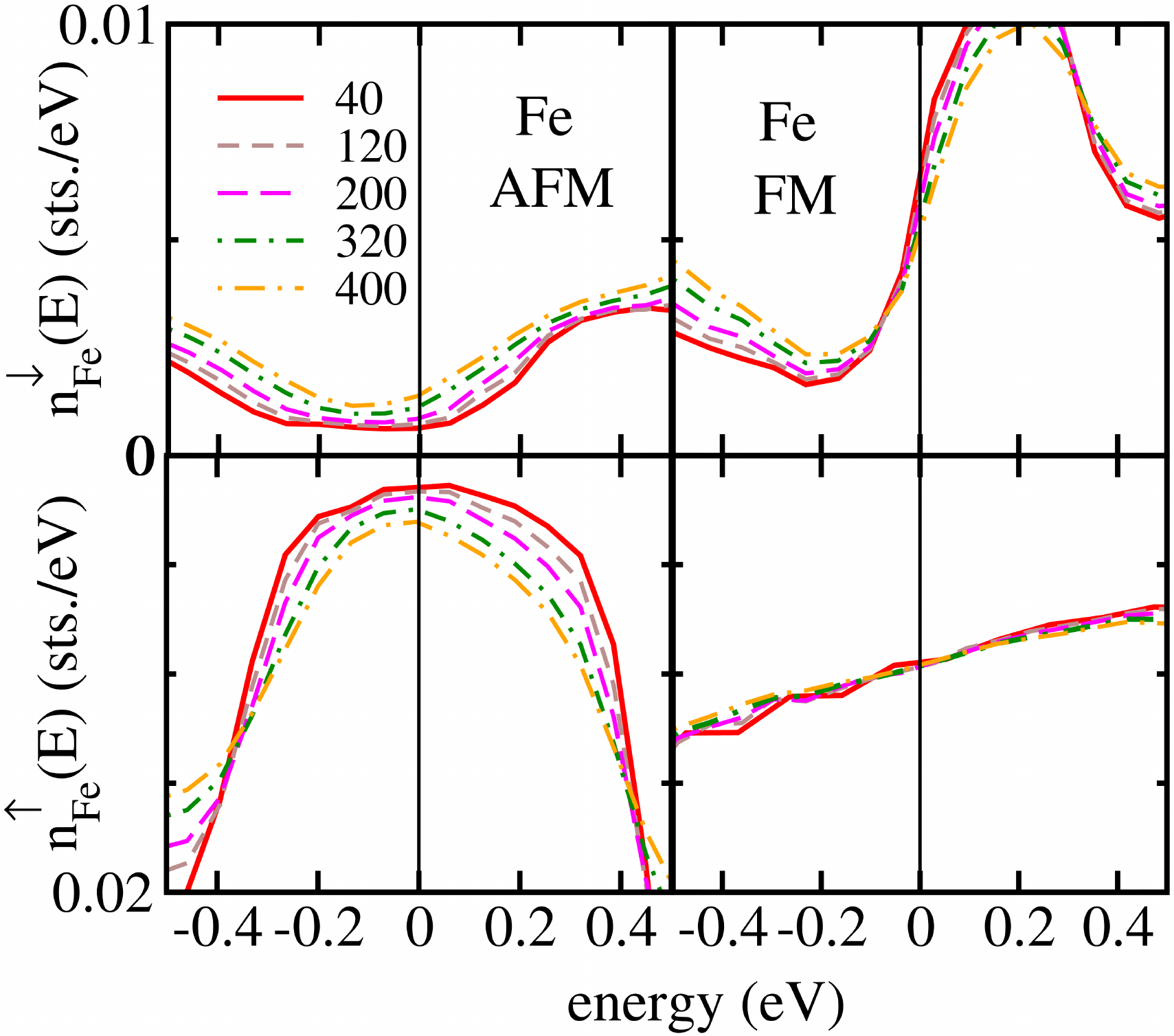}
 \includegraphics[width=0.23\textwidth,angle=0,clip]{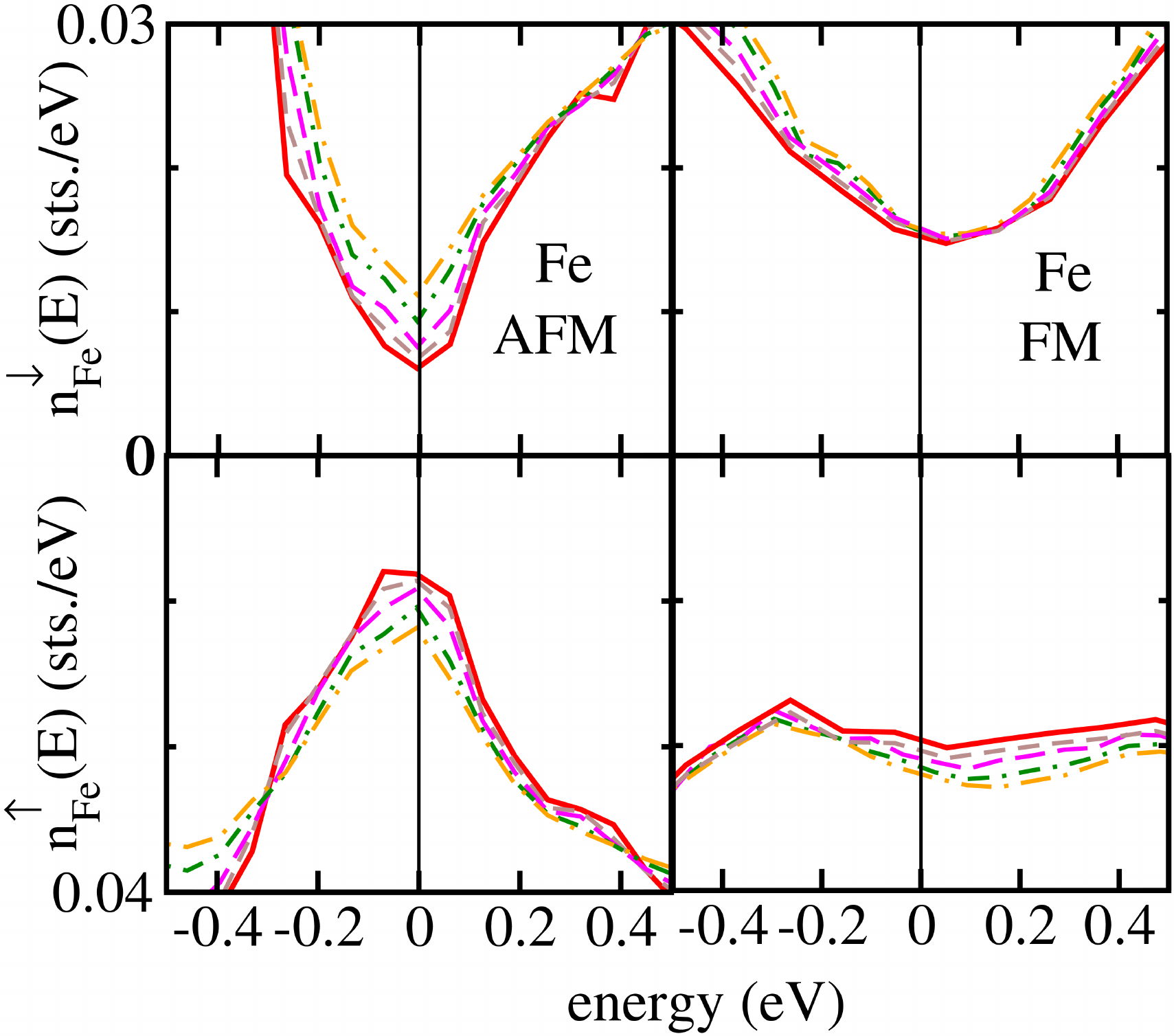}
 \\ (a)  \;\; \;\; \;\; \;\; \;\; \;\; \;\; \;\; \;\; \;\;  \;\; \;\;       (b)\\
 \includegraphics[width=0.23\textwidth,angle=0,clip]{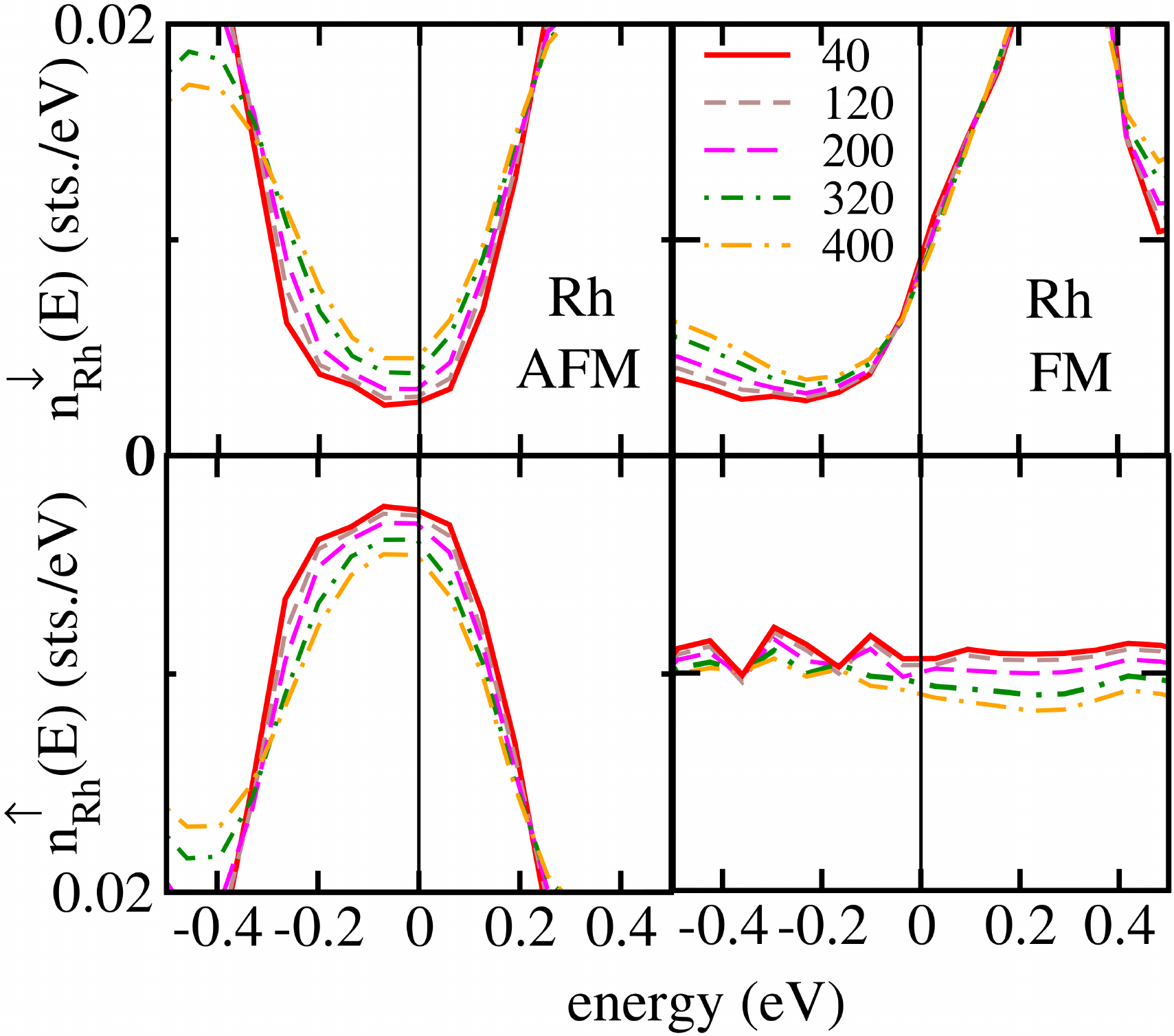}
 \includegraphics[width=0.23\textwidth,angle=0,clip]{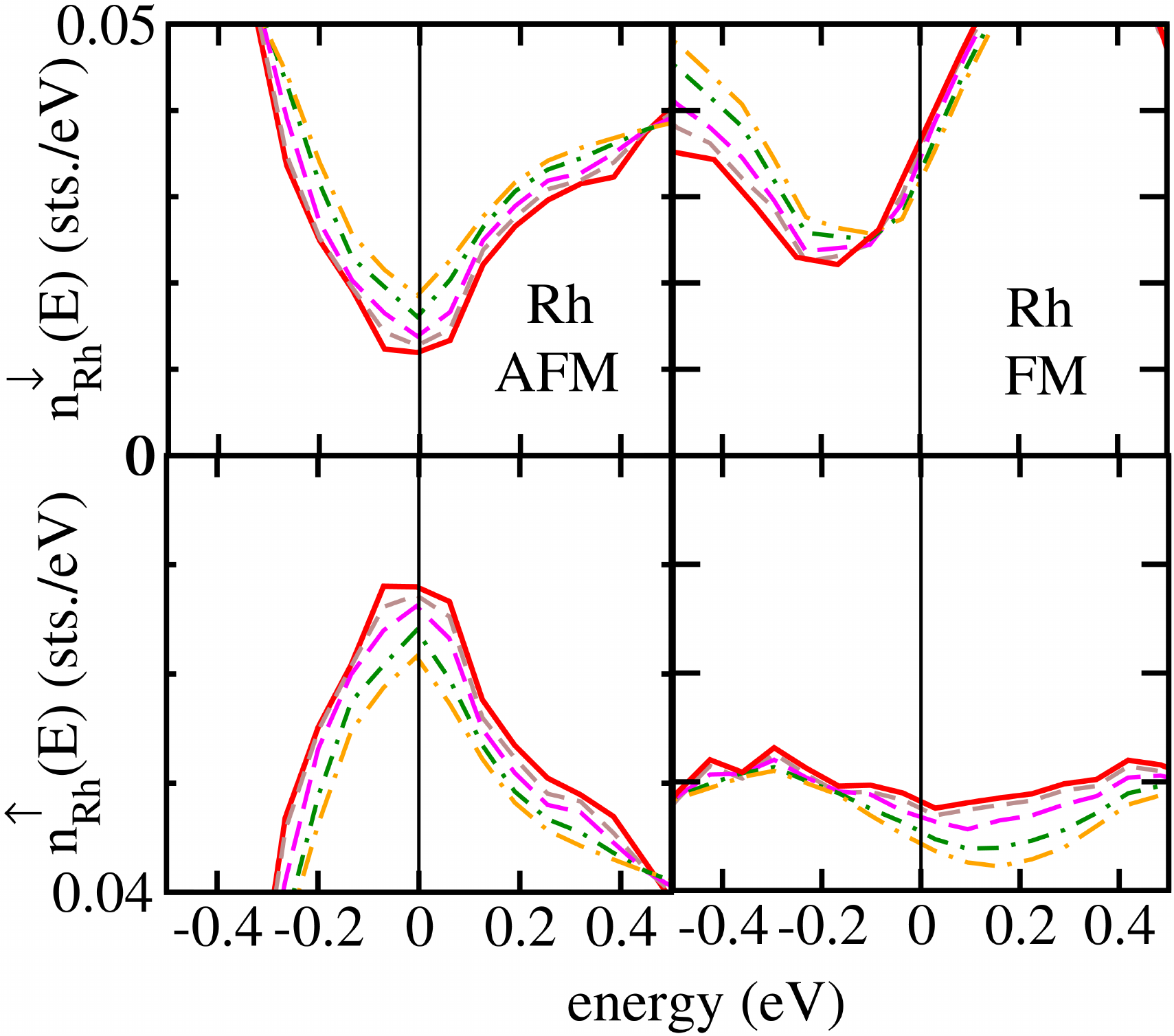}
 \\(c)     \;\; \;\; \;\; \;\; \;\; \;\; \;\; \;\; \;\; \;\;  \;\; \;\;     (d) 
 \caption{\label{fig:sp-DOS_0} Comparison of the temperature dependent
   densities of states (DOS) for the FM and AFM states of FeRh for $T =
   40--400$~K : (a) Fe $s$-DOS, (b) Fe $p$-DOS, (c) Rh $s$-DOS, and (d) Rh $p$-DOS.
  }  
 \end{figure}

We use the calculated density of states at the Fermi level as a measure of
the concentration of the conducting electrons.  The change of the carriers
concentration at the AFM-FM transition can therefore be seen from the
modification of the $sp$-DOS at the Fermi level. The element-projected
spin-resolved $sp$-DOS ($n_{sp}(E)$), calculated for both FM and AFM
states at different temperatures, is shown in
Fig.\ \ref{fig:sp-DOS_0}. At low temperature, for both 
Fe and Rh sublattices, the $sp$-DOS at $E_F$ is higher in the FM than in
the AFM state,
$n^{FM}_{sp}(E_F) > n^{AFM}_{sp}(E_F)$. This gives a first hint concerning
the origin of the large difference between the FM- and AFM-conductivities in the
low temperature limit (see inset for $\sigma^{vib}_{xx}$ in
Fig. \ref{fig:FeRh_Rho}(b)). In this case the relaxation time $\tau$ is
still long owing to the low level of both lattice vibrations and spin
fluctuations which determines the scattering  
potential $V_{scatt}$. For both magnetic states the decrease of the conductivity with rising  
temperature is caused by the increase of
scattering processes and consequent decrease of the relaxation
time. At the same time, the conductivity difference, 
$\Delta\sigma(T) = \sigma^{vib,FM}_{xx}(T) - \sigma^{vib,AFM}_{xx}(T)$, reduces with increase in
temperature. This effect can partially be attributed to the
temperature dependent changes of the electronic structure (disorder smearing of
the electronic states) reflected by changes in the density of states at
the Fermi level~\cite{SBS+14} (see Fig. \ref{fig:sp-DOS_0}). Despite this, up to the
transition temperature, $T = T_m$, the difference $\Delta\sigma(T)$ is
rather pronounced leading to a significant change of the
resistivity at  $T = T_m$. 

\begin{figure}[h]
\includegraphics[width=0.29\textwidth,angle=0,clip]{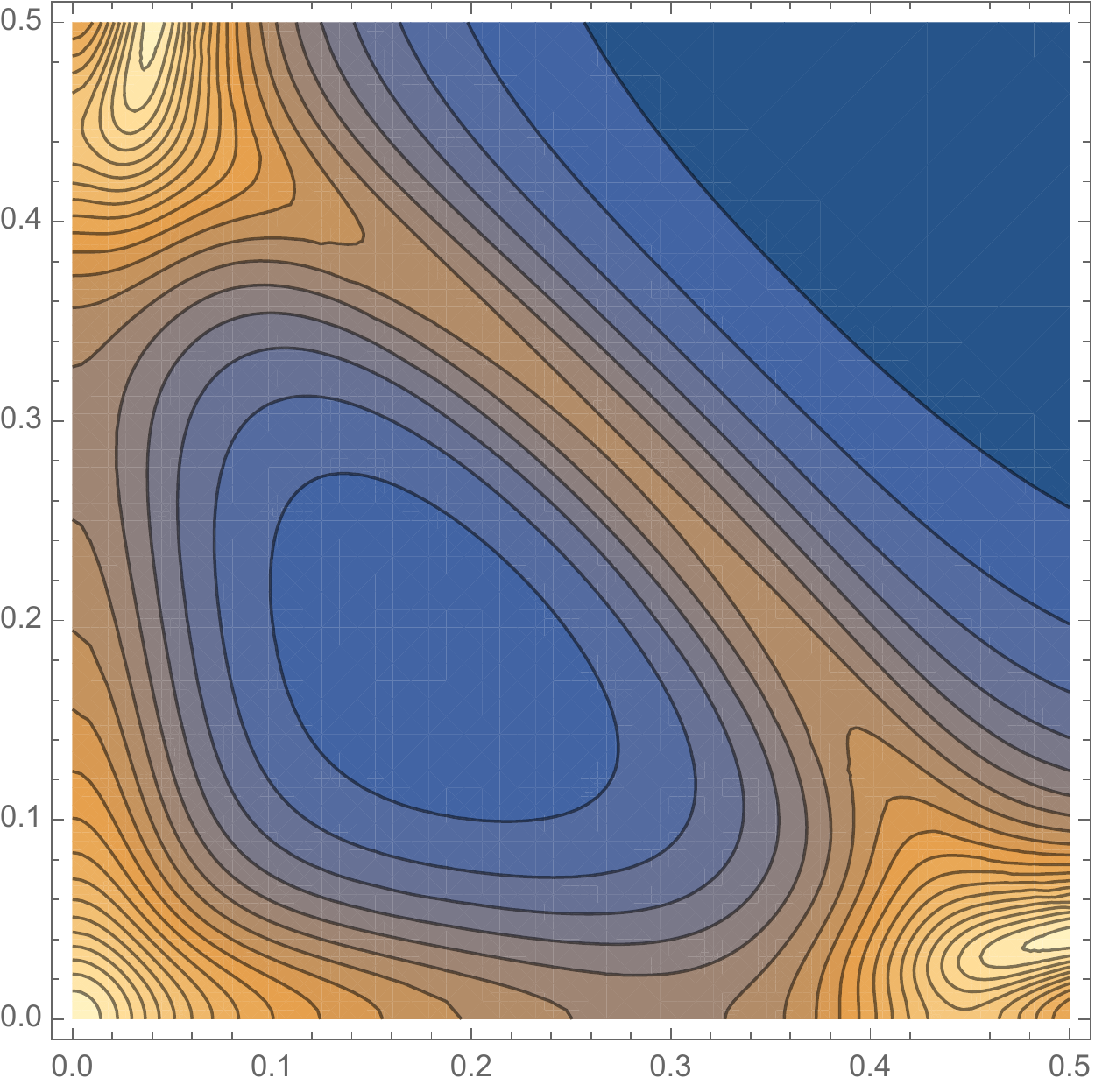}\;(a)\\
\includegraphics[width=0.29\textwidth,angle=0,clip]{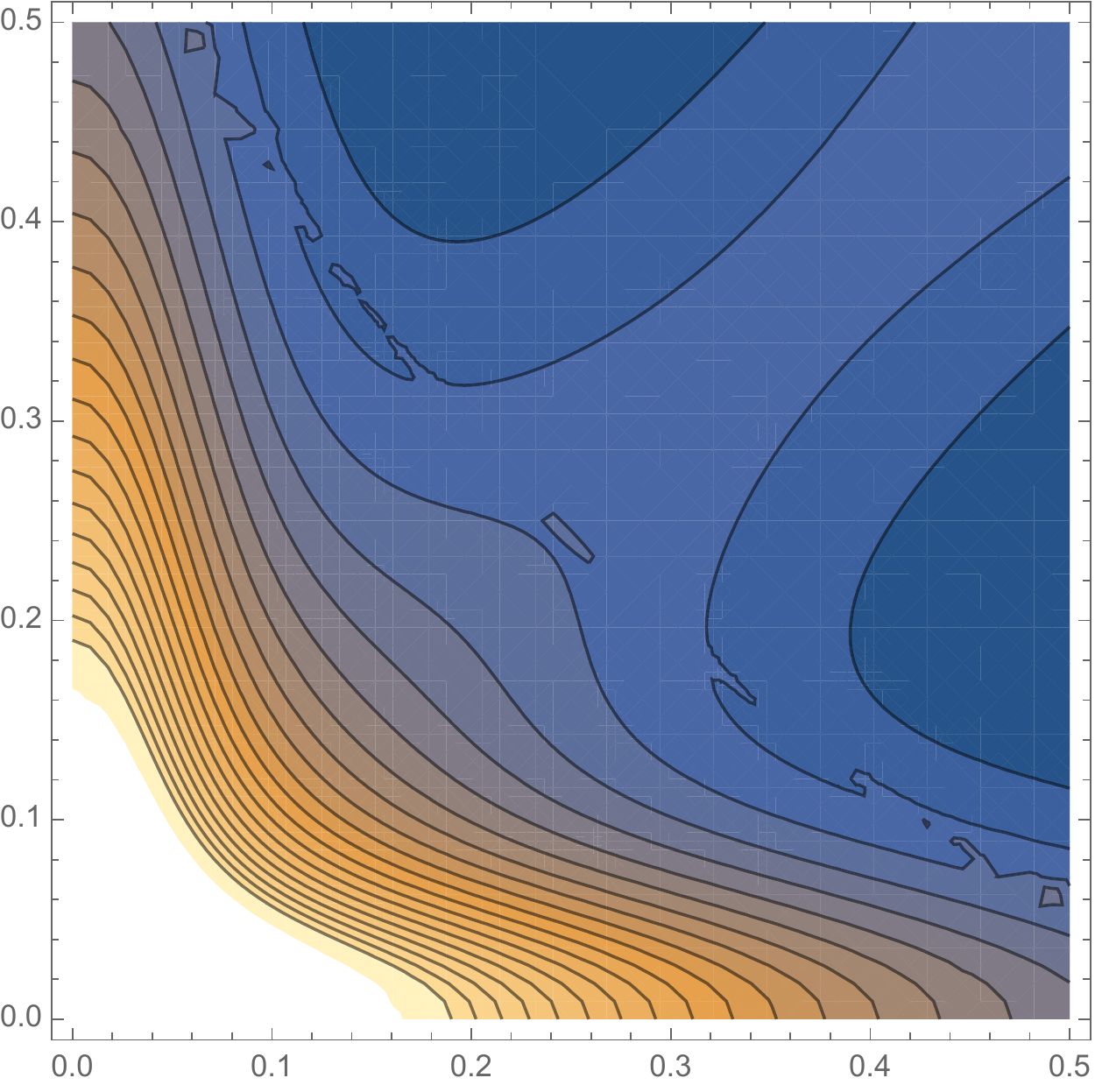}\;(b)\\ 
\includegraphics[width=0.29\textwidth,angle=0,clip]{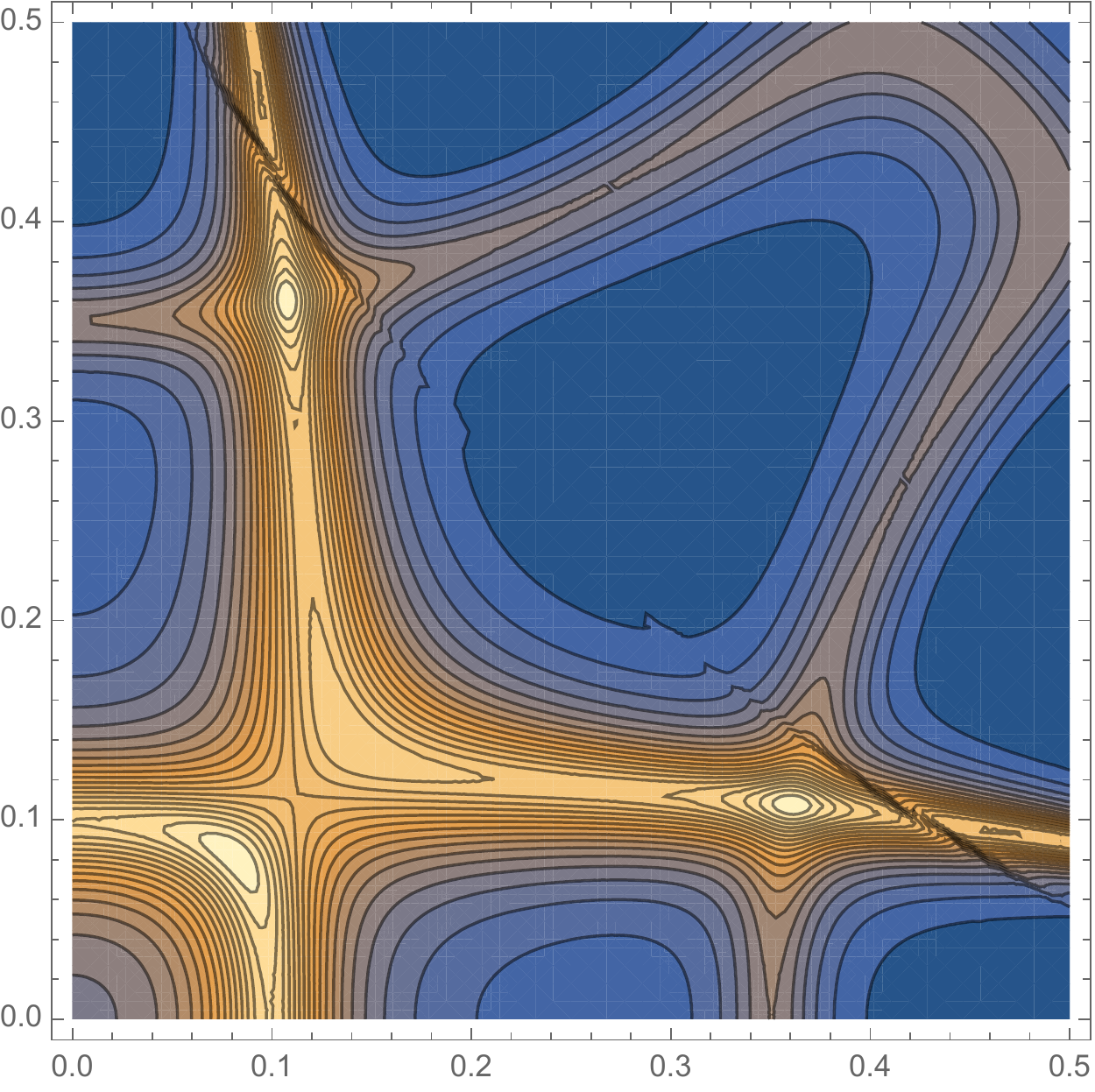}\;\;(c)
\caption{\label{fig:FeRh_BSF} (a) Bloch spectral function of FeRh
  calculated for the AFM state at $T = 300$~K (a) and for the FM state
  resolved into majority spin (b) and minority spin (c) electron
  components, calculated for $T = 320$~K. The finite width of this
  features determine the electronic mean free paths.  }  
\end{figure}

One has to stress that in calculating the contribution of spin moment
fluctuations to the resistivity, the different 
temperature dependent behavior of the magnetic order in the FM and AFM states must be taken into
account. This means, that at the critical point, $T = T_m$, the smaller
sublattice magnetization in the AFM state describes a more pronounced
magnetic disorder when compared to the FM state which leads to both a smaller
relaxation time and shorter mean free path. The result is a
higher resistivity in the AFM state.

The different mean free path lengths in the FM and AFM states at a given
temperature can be analyzed using the Bloch spectral function (BSF),
$A_B(\vec{k}, E)$ \cite{SM}, calculated for $E = E_F$, 
since the electronic states
at the Fermi level give the contribution to the electrical conductivity.
 For a system with thermally induced spin fluctuations and lattice
displacements the BSF has features with 
finite width from which the mean free path length of the electrons can be
inferred. Fig.~\ref{fig:FeRh_BSF} shows an intensity contour  
plot for the BSF of FeRh averaged over local moment configurations
appropriate for the FM and AFM states just above and just below the FM-AFM transition
respectively. Fig.~\ref{fig:FeRh_BSF}(a) shows the AFM Bloch spectral 
function whereas Figs.~\ref{fig:FeRh_BSF}(b) and (c) show the sharper
features of the spin-polarized BSF of the FM state especially
for the minority spin states. This implies a longer electronic mean free
path in the FM state in comparison to that in the AFM state which is 
consistent with the drop in resistivity. 

\begin{figure}[h]
\includegraphics[width=0.4\textwidth,angle=0,clip]{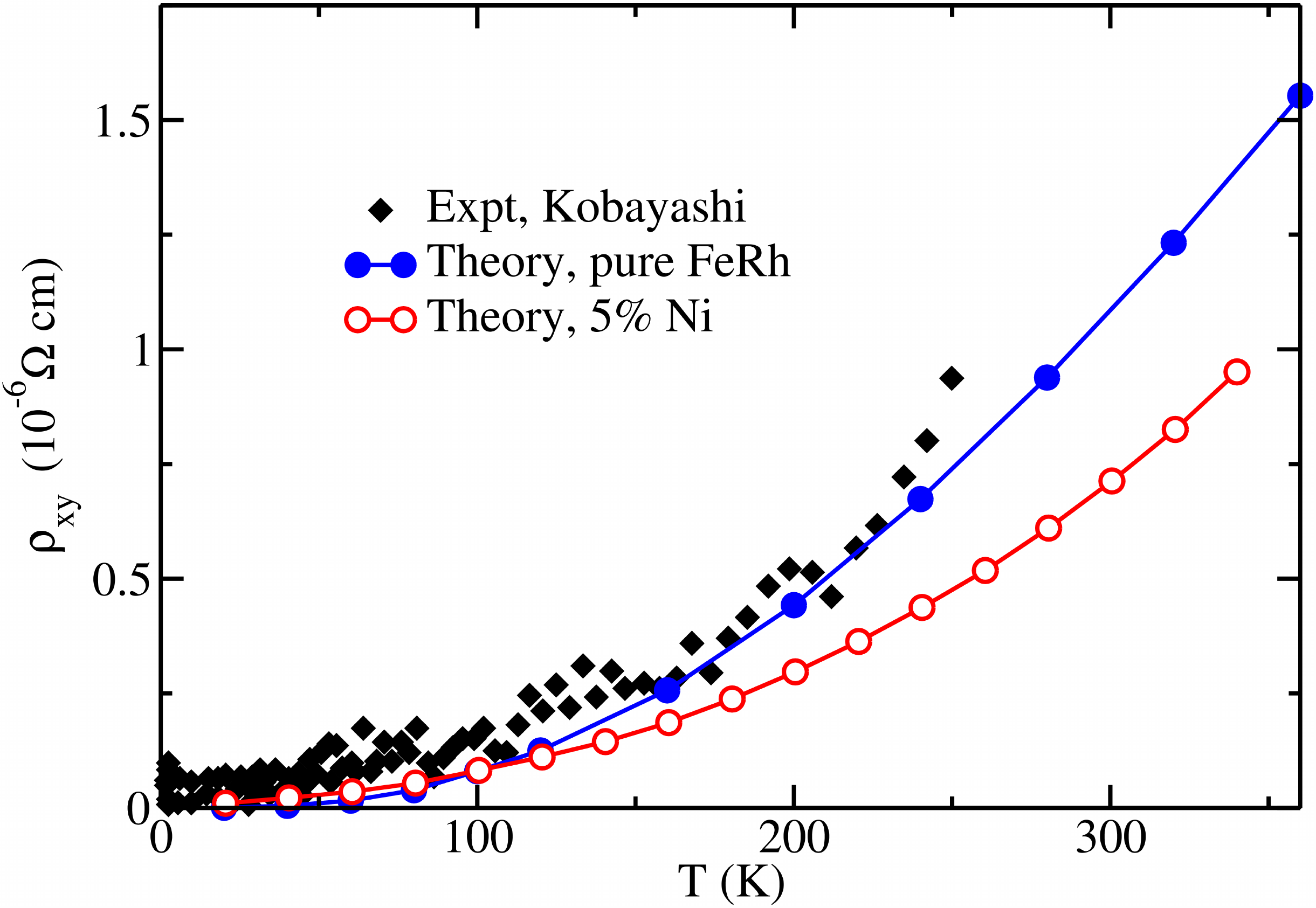}\;(a) \\
\includegraphics[width=0.4\textwidth,angle=0,clip]{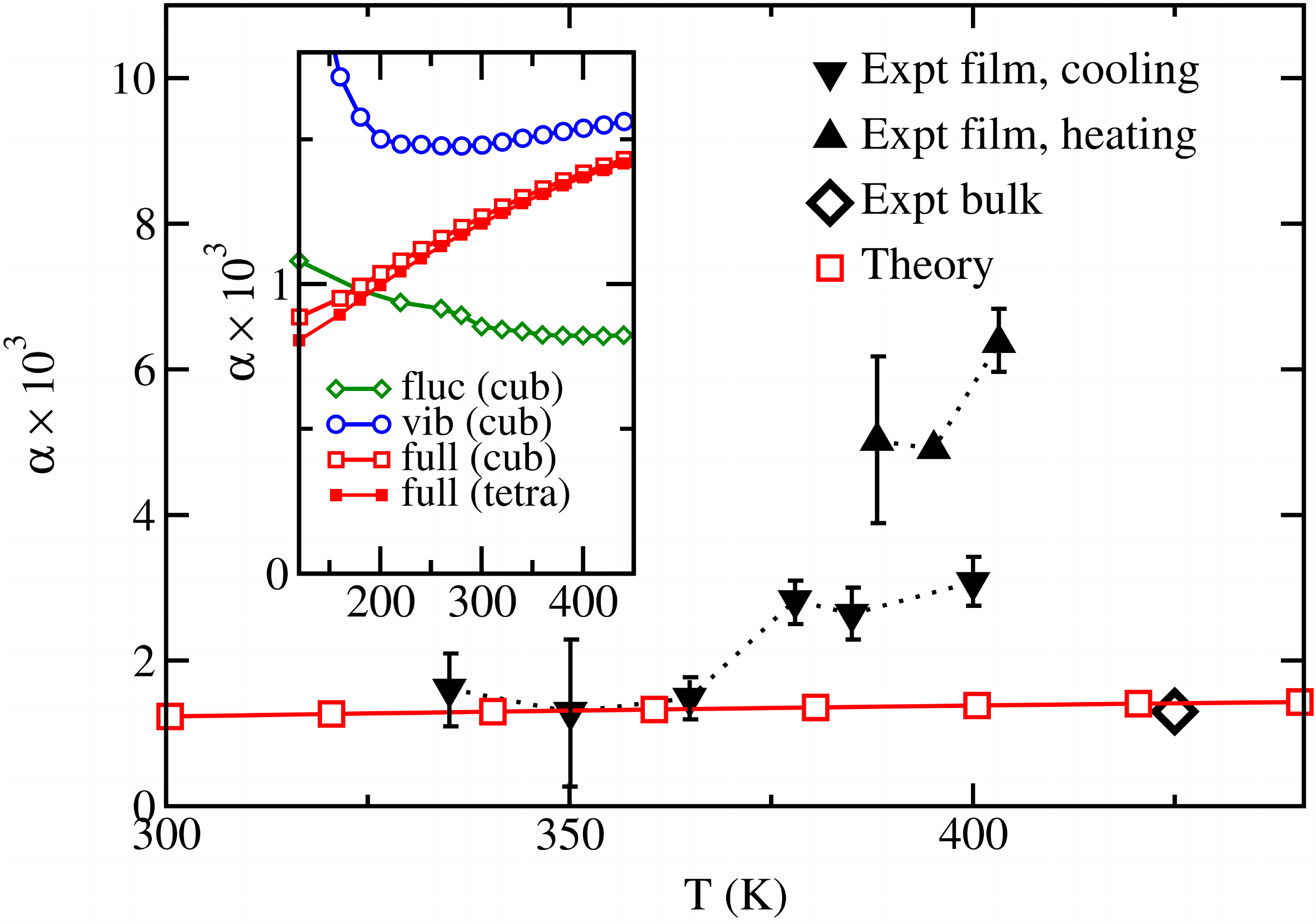}\;(b)
\caption{\label{fig:FeRh_AHE_Rho} (a) The temperature dependence of the
  anomalous Hall resistivity for the FM state of (Fe$_{0.95}$Ni$_{0.05}$)Rh in
  comparison with experimental data \cite{KMA01}; 
  (b) Gilbert damping parameter as a function of
  temperature: theory accounting for all thermal contributions (squares)
  in comparison with the experimental results for thick-film system (50~nm)
  \cite{MPH+13} (open diamond) and for 
  FeRh thin film deposited on MgO(001) surface (up- and down-triangles).
  Up- and down-triangles represent data for a heating and cooling
  cycles, respectively (for details see supplementary materials).
  The inset represents the results for the individual sources for the
  Gilbert damping, i.e., lattice vibrations (circles) and spin fluctuations
  (diamonds). The total $\alpha$ values calculated for FeRh crystal
  without (c) and with  tetragonal (t) distortions ($c/a = 1.016$) are shown by
  open and closed squares, respectively. 
 }  
\end{figure}
In particular concerning technical applications of FeRh, it is
interesting to study further temperature dependent response properties.
In Fig. \ref{fig:FeRh_AHE_Rho}(a) we show our calculations of the 
 total anomalous Hall resistivity for FeRh in the FM state, represented by
the off-diagonal term $\rho_{xy}$ of the resistivity tensor and compare it 
with experimental data  \cite{KMA01}.
As the FM state is unstable in pure FeRh at low temperatures, the
measurements were performed for (Fe$_{0.965}$Ni$_{0.035}$)Rh, for which
the FM state has been stabilized by Ni doping.
The calculations have been performed both, for the pure FeRh compound as well
as for FeRh with $5\%$ Ni doping,  (Fe$_{0.95}$Ni$_{0.05}$)Rh, which theory finds
to be ferromagnetically ordered down to $T=$0~K. As can be seen the magnitude
of $\rho_{xy} (T)$ increases in a more 
pronounced way for the undoped system. Nevertheless, both results are
in a rather good agreement with experiment. 

In addition to the temperature dependent transport properties 
the inclusion of relativistic effects into the ab-initio theory enables us to 
present results for the Gilbert damping, which plays a crucial
role for spin dynamics. We have calculated this quantity taking into
account all temperature induced effects, i.e. spin fluctuations and 
lattice vibrations \cite{MKWE13, EMC+15}. As one can see in Fig. \ref{fig:FeRh_AHE_Rho}(b),
the calculated results are in rather good agreement with the
experimental value (shown by diamond) $\alpha = 0.0012$ obtained for a
thick film at $T = 420$~K \cite{MPH+13} as well as new experimental data
for thin films \cite{SM}.  
The separate contributions to the Gilbert damping due to spin fluctuations
and lattice vibrations are presented in the inset to
Fig. \ref{fig:FeRh_AHE_Rho}(b) for a given temperature window again
artificially extended to low temperatures. These results allow to
identify the leading role of lattice vibrations (circles in the inset
to Fig. \ref{fig:FeRh_AHE_Rho}(b)) at high temperature 
region where the electron spin-flip interband transitions are most
responsible for dissipation due to the magnetization dynamics. In the
low-temperature 
region, where the $T$-dependence of $\alpha$ is determined by intraband
spin-conserving scattering events, it
stems dominantly from 
electron scattering due to thermally induced spin-fluctuations
(diamonds in the inset to Fig. \ref{fig:FeRh_AHE_Rho}(b)).

The experimental data shown in Fig. \ref{fig:FeRh_AHE_Rho}(b)) by 
 triangles represent results for rather thin films 
 ($d = 25$~nm) deposited on top of a MgO(001) substrate \cite{SM}.  
The FeRh unit cell with a lattice constant $~\sqrt{2}$ times smaller
than that of MgO, is rotated around $z$ axis by $45^o$
with respect to the MgO cell. Because of this,
 a compressive strain in the FeRh
film occurs. As it follows from the experimental data
\cite{BJC+12}, this implies a tetragonal distortion of the FM FeRh unit
cell with $c/a = 1.016$. Results of  corresponding  calculations
for $\alpha$
are given in the inset of Fig. \ref{fig:FeRh_AHE_Rho}(b) by full
squares, demonstrating a rather weak effect of this distortion. 
The smaller value of  $\alpha$ compared to experiment,
has therfore to be attributed to the use of bulk geometry
instead of the experimental film geometry with a corresponding
impact on the damping parameter.
\medskip

In summary, we have presented ab-initio calculations for the finite
temperature  transport properties of the FeRh compound. A steep increase of
the electric resistivity has been obtained for the AFM state leading to
a pronounced drop of resistivity at the AFM to FM transition temperature. 
This effect can be attributed partially to the difference of the
electronic structure of FeRh in the FM and AFM states, as well as
to a faster increase of the amplitude of spin fluctuations caused by
temperature in the AFM state.  Further calculated temperature dependent
response properties such as the AHE resistivity and the Gilbert damping
parameter for the FM system show also good agreement with
experimental data.  This gives additional confidence  in the model used
to account for thermal lattice vibrations and spin fluctuations.

\section{Acknowledgements}

Financial support by the DFG via SFB 689 (Spinph\"anomene in reduzierten
Dimensionen) and from the EPSRC (UK) (Grant No. EP/J006750/1) is gratefully acknowledged.


%

\end{document}